\documentclass[showpacs,prb,twocolumn,aps,superscriptaddress,floatfix,amsmath,amssymb]{revtex4-1}

\usepackage{color,graphicx,bm}

\newcommand{\rhos}{\ensuremath{\rho_s}}
\newcommand{\heff}{\ensuremath{H_{\mathrm{eff}}}}

\newcommand{\ld}{\ensuremath{L_{\mathrm{d}}}}
\newcommand{\Id}{\ensuremath{I_{\mathrm{d}}}}
\newcommand{\jcf}{\mbox{${\bm{j}}_{\text{cf}}$}}
\newcommand{\jd}{\mbox{${\bm{j}}_{\text{d}}$}}
\newcommand{\jp}{\mbox{${\bm{j}}_{\text{p}}$}}
\newcommand{\bnabla}{\mbox{${\bm{\nabla}}$}}
\newcommand{\xuvec}{\mbox{${\bm{\hat{x}}}$}}
\newcommand{\bq}{\mbox{${\bm{q}}$}}
\newcommand{\qdotr}{\mbox{${\bm{q}} \cdot {\bm{r}}$}}
\newcommand{\Mpara}{M_\parallel}
\newcommand{\Mparamax}{M_{\parallel,\mathrm{max}}}
\newcommand{\Mperp}{M_{\perp}}

\newcommand{\derek}[1]{$\spadesuit${\sl #1}}
\renewcommand{\derek}[1]{}

\begin{document}

\title{Diamagnetism and flux creep in bilayer exciton superfluids}

\author{P. R. Eastham}\affiliation{School of Physics, Trinity College, Dublin 2, Ireland.}

\author{N. R. Cooper}\affiliation{T.C.M. Group, Cavendish Laboratory, 
J.J. Thomson Avenue, Cambridge CB3 0HE, United Kingdom}

\author{D. K. K. Lee}\affiliation{Blackett Laboratory, Imperial
College London, London SW7 2AZ, United Kingdom}

\date{\today}

\begin{abstract}
  We discuss the diamagnetism induced in an isolated quantum Hall
  bilayer with total filling factor $\nu_T=1$ by an in-plane magnetic
  field. This is a signature of counterflow superfluidity in these
  systems. We calculate magnetically induced currents in the presence
  of pinned vortices nucleated by charge disorder, and predict a
  history-dependent diamagnetism that could persist on laboratory
  timescales. For current samples we find that the maximum in-plane
  moment is small, but with stronger tunneling the moments would be
  measurable using torque magnetometry. Such experiments would allow
  the persistent currents of a counterflow superfluid to be observed
  in an electrically isolated bilayer.
\end{abstract}

\maketitle

\section{Introduction}

Superfluidity\cite{Leggett1999} is a spectacular form of hydrodynamics
involving dissipationless flow, metastable circulation, and
quantization of circulation. It occurs in liquid helium and cold
atomic gases, where it is associated with the condensation of many
bosons into a single quantum state. Whether condensates of
quasiparticles, such as excitons, would also be superfluids has been
discussed for many years, with debate over the physical
manifestations\cite{Blatt1962,Kohn1970,Hanamurat1974,Lozovik1975,BalatskyA.V.2004,Su2008,Bunkov2010}
of superfluid hydrodynamics for quasiparticles, the (related) role of
symmetry-breaking perturbations, and the significance of interactions
and thermal equilibrium\cite{Wouters2010}.

\begin{figure}[hbt]
\includegraphics{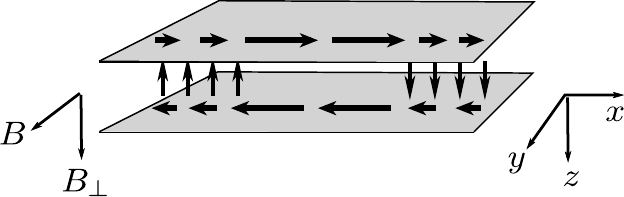}
\caption{Schematic diagram of a bilayer of electrons in the quantum
  Hall regime due to a strong perpendicular field $B_\perp$.  Applying
  a weak in-plane field $B$ causes counterflowing charge currents as a
  diamagnetic response. For an isolated bilayer the current loop
  closes at the edges of the sample by interlayer tunneling.  In an
  excitonic picture, this corresponds to the flow of neutral excitons
  along the bilayer with recombination at the
  edges.}\label{fig:bilayer}
\end{figure}

This question can now be addressed experimentally, with emerging
evidence for the condensation of quasiparticles including excitons,
polaritons, and magnons. A particularly interesting system is the
quantum Hall bilayer at total filling factor $\nu_T=1$ (see
Fig.~\ref{fig:bilayer}). This consists of two closely spaced quantum
wells, each containing a two-dimensional electron gas, subjected to a
strong perpendicular field $B_\perp$ so that the lowest Landau level
in each layer is half-filled. The tunneling between the two layers is
weak compared to the Coulomb energy scale.  When the interlayer
separation $d$ is of the order of the magnetic length
$l_B=(\hbar/eB_\perp)^{1/2}$, the ground state is believed to be a
Bose-Einstein condensate of interlayer
excitons\cite{Eisenstein2004,Fertig1989,Wen1993,Ezawa1993,Murphy1994,Lay1994}
caused by the Coulomb attraction of electrons and holes across the
layers. (Note that the holes are unfilled electron states of the
lowest Landau level of the conduction band.)  Flows of these condensed
excitons correspond to dissipationless counterflowing electrical
currents in the two layers\cite{Wen1993}.  Initial evidence of
dissipationless transport came from interlayer
tunneling,\cite{Spielman2000,Finck2008} which exhibits non-zero
interlayer currents at negligible interlayer
voltages\cite{Yoon2010,Tiemann2009,Tiemann2008,Wen1993}. This regime
persists up to a critical current, above which dissipation increases
rapidly.

The interpretation of transport measurements as
evidence\cite{Snoke2011} for exciton superfluidity is complicated by
parallel charge transport channels, the injection and removal of
electrons\cite{Su2008,Su2010,Pesin2011} at the contacts, as well as
possible dissipation in the leads. Here, we show that magnetometry on
\emph{isolated} bilayers could provide direct evidence for exciton
superfluidity, without the complications inherent in transport
studies. We predict that, at low temperatures, the bilayer shows a
history-dependent susceptibility. Changes in the in-plane field lead
to a persistent diamagnetic moment that is, however, not induced by
fields present when the condensate forms. This discrepancy corresponds
to that between the moments of inertia of normal and superfluid helium
inferred in a torsional oscillator experiment.

The in-plane magnetic susceptibility of the exciton condensate has
previously been
considered\cite{Hanna2001,Yang1994,Shevchenko1997,BalatskyA.V.2004} in
clean bilayers.  However, there are very good reasons to
expect\cite{Lee2011,Eastham2010,Fogler2001,Fertig2005} that disorder
is essential for understanding the observed transport properties of
the bilayer at low temperatures.  In particular, for quantum Hall
bilayers, charge disorder nucleates vortices in the
condensate\cite{Eastham2009a,Fertig2005,Stern2001,Fogler2001,Balents2001}.
Therefore, we will study in this work the magnetic properties of a
quantum Hall bilayer with disorder-induced vortices. We first consider
the zero-temperature limit, and show how dissipationless supercurrents
lead to long-lived diamagnetic moments. In current samples, the
resulting moments are small [Eq.~\eqref{eq:maxmoment}], but for
samples with stronger tunneling the moments can be comparable to the
Landau diamagnetism. We also consider the effects of non-zero
temperatures, and estimate the dissipation due to thermal
creep\cite{ANDERSON1964} of the in-plane flux.  This dissipative
mechanism gives rise to a non-vanishing resistance for the counterflow
supercurrents, and causes the diamagnetic moments to decay in
time. Based on the parameters of current experiments, we expect
dissipation to be significant, which is consistent with experimental
resistance measurements. The flux motion in this regime can be studied
in transport experiments. Samples deeper in the condensed phase,
however, would show persistent counterflow supercurrents on laboratory
timescales, whose presence and eventual decay could be observed by
magnetometry.

The remainder of this paper is structured as follows. In
Sec. \ref{sec:model}, we outline the model we consider for the behavior
of long-wavelength counterflow supercurrents in the disordered
bilayer. In Sec. \ref{sec:zerot}, we apply this model to calculate the
in-plane susceptibility of the bilayer in the zero-temperature limit. In Sec. \ref{sec:finitet}, we discuss the extension to finite but
low temperatures. Finally, in Secs. \ref{sec:discuss} and
\ref{sec:conclusions}, we present numerical estimates for the moments
and resistance in experiments, and summarize our conclusions.

\section{Model and background}
\label{sec:model}

In this work, we consider a quantum Hall bilayer at total filling
factor $\nu_T=1$, in the presence of the disordered electrostatic
potential originating from the dopants, set back at a distance $d_d$.
For definiteness, we adopt the ``coherence network'' picture of Fertig and
Murthy,\cite{Fertig2005} although our theory here is more general. In
this picture, the random Coulomb field from the dopants creates
puddles of normal electron liquid surrounded by channels of excitonic
superfluid. In quantum Hall ferromagnets, physical and topological
charge are related so that the charge puddles nucleate vortices, known
as merons, in the superfluid
channels\cite{Eastham2009a,Fertig2005,Stern2001,Fogler2001,Balents2001}.
We estimated\cite{Eastham2009a} that typical disorder strengths induce
on the order of one unpaired vortex per puddle so that the correlation
length of this disorder is $\xi\approx d_d \approx$ 100 nm.

The energy associated with the excitonic supercurrents in the channels
is, in terms of a superfluid phase $\eta$,\cite{Eastham2010}
\begin{equation}
  \heff =\int \left[
  \frac{\rhos}{2} |\bnabla \eta+ \bq|^2 - t \cos (\eta+\theta_0)
  \right]d^{2}r, 
\label{heff-etagauge}
\end{equation}
where the first term is the kinetic energy of the counterflow
supercurrent with superfluid stiffness $\rhos$, and the second term is
the energy of the interlayer tunneling currents with tunneling
strength $t$. The merons introduced by the charge disorder give rise
to the random field $\theta_0$. For the coherence network, both
$\rhos$ and $t$ should be renormalized by the area fraction of the
superfluid channels.

Eq. (\ref{heff-etagauge}) is written in a gauge where the vector
potential of the in-plane field $B$ is zero perpendicular to the
layers and non-zero parallel to the layers. This field then appears in
the kinetic energy, inducing a
wavevector\cite{Wen1993,BalatskyA.V.2004,Rademaker2011}
\begin{equation}
\bq=({\bm{B}} \times \hat{ {\bm{z}}}) e d/\hbar\,,
\label{eq:qdef}
\end{equation} 
where $d$ is the interlayer separation. The counterflow and tunneling
currents are seen to be
\begin{eqnarray}
  \jcf&=&\frac{e\rhos}{\hbar} (\bnabla \eta+\bq)=\jp + \jd, 
\label{eq:cfcurrent} \\ 
  j_t&=&\frac{e t}{\hbar}\sin (\eta+\theta_0),
\label{eq:tcurrent} 
\end{eqnarray} 
so the magnetic field induces a diamagnetic contribution $\jd= e\rhos
\bq/\hbar$ to the counterflow supercurrent $\jcf$, but does not appear
explicitly in the coherent tunneling current $j_{t}$.  We can make a
gauge transformation so that the vector potential is non-zero only
perpendicular to the bilayer, in which case the phase transforms to
$\eta\to \eta+\qdotr\equiv\theta$:
\begin{equation} H_{\rm eff} =\int
  \left[\frac{\rhos}{2} |\bnabla \theta|^2 
  - t \cos (\theta-\qdotr+\theta_0)
  \right]d^{2}r. 
\label{heff-thetagauge} 
\end{equation} This form for the energy functional highlights the fact that the
in-plane magnetic response of the bilayer vanishes if there are no current
loops that encircle the field --- the field disappears from
Eq.~\eqref{heff-thetagauge} when $t=0$.  (A non-zero diamagnetic
susceptibility remains possible in multiply-connected
geometries\cite{Rademaker2011} or at non-zero
frequency\cite{BalatskyA.V.2004}. We comment on the extension of our
results to the Corbino disk geometry in Sec.\ref{sec:discuss}.)

At zero temperature, the state of the bilayer is determined by
minimizing Eq.~\eqref{heff-etagauge}, which gives
\begin{equation} 
-\lambda^2_J\nabla^2\eta +\sin(\eta+\theta_0)=0\,,\qquad
\lambda_J = (\rhos/t)^{1/2}\,.
\label{etapendulum}
\end{equation} When $\theta_0=0$ this is the pendulum equation, containing the Josephson length $\lambda_J$: the characteristic lengthscale over which excitonic supercurrents decay by interlayer tunneling. This scale is estimated to be on the order of a few microns in experiments. We will discuss below the decay of excitonic supercurrents by tunneling in the disordered case, which involves a different lengthscale.

It is useful to recall, for comparison with our treatment of the
disordered case, the
response\cite{Bak1982,Hanna2001,Fil2010,Read1995,Yang1994} of the
clean model ($\theta_0=0$) to an in-plane field. For small fields, the
ground state will minimize the tunneling energy, and so
$\theta\approx\qdotr$ or $\eta\approx 0$. This is the commensurate
state, in which the field induces a long-wavelength counterflow
supercurrent $\jcf\approx \jd$, as in Eq.~\eqref{eq:cfcurrent}. Thus
there is an in-plane magnetic moment of $\Mpara = j_d d L_x L_y = d
L_x L_y\chi_0 B/\mu_0$, with the susceptibility
\begin{equation} 
\chi_0=\mu_0 j_d/B=\mu_0 e^2\rhos d/\hbar^2\,.
\label{chi0}
\end{equation} We see that the diamagnetic moment $\Mpara = \chi_0 B$ increases linearly
with the in-plane field $B$ in the commensurate state.

In an isolated bilayer, the counterflow currents must vanish at the
ends of the sample. This occurs in the commensurate phase because, as
dictated by Eq. (\ref{etapendulum}), the diamagnetic currents in the
bulk of the sample are eliminated by tunneling over a region of size
$\lambda_J$ near the sample boundaries
(Fig.~\ref{fig:bilayer}). However, since the phase $\eta$ is a
periodic variable the maximum current which can recombine in this way
is given by $e\rho_s|\bnabla\eta|/\hbar\sim
e\rho_s/\hbar\lambda_J$. If the diamagnetic current present in the
bulk exceeds this value, then a net winding of the phase $\eta$, in the
form of Josephson vortices, enters from the boundaries. This occurs at
the field \begin{equation} B_0\sim \hbar/ed\lambda_J\,.
  \label{eq:b0}
\end{equation}  Above $B_0$ the system is in the incommensurate phase
in which the kinetic energy of the counterflow supercurrents dominates
so that $\theta\approx 0$ or $\eta \approx -\qdotr$. The net winding
in $\eta$ along the sample implies that the counterflow supercurrents
have weak oscillations around zero along the sample, and the
diamagnetic moment is small. The field $B_0$ marks the field above which the diamagnetic susceptibility rapidly decreases from
$\chi_0$ as Josephson vortices fill the system and compensate the
diamagnetic contribution in Eq.~\eqref{eq:cfcurrent}.

While in the clean limit the phase twists nucleated at the boundary
propagate into the bulk, in the disordered case any such phase twists
can be pinned by the disorder. They can thus be kept out of the bulk
of the sample, which can continue to contribute the full diamagnetic
susceptibility $\chi_0$. To describe this effect quantitatively we
shall use the collective pinning theory we have previously
applied\cite{Lee2011,Eastham2010} to the transport experiments in zero
field ($q=0$).

The starting point for understanding the transport experiments is to
note that Eq.~(\ref{heff-etagauge}) is a random-field XY model. The
energy $\heff$ consists of the competition between the tunneling
energy, which is minimized by a spatially varying superfluid phase
$\eta$ over the disorder correlation length $\xi$, and the superfluid
stiffness, which is minimized by a uniform superfluid phase. For the
bilayer, the disorder correlation length is much shorter than the
clean tunneling length: $\xi \ll \lambda_J$.  In this regime, the
superfluid stiffness dominates at short scales, up to the Imry-Ma or
pinning length
\begin{equation}
\ld\sim \lambda_J^2/\xi = \rhos/t\xi\,
\label{eq:domainsize}
\end{equation} where the two energies balance. We
estimate\cite{Eastham2010} that $\ld\approx$ 10 -- 100 $\mu$m and
$\rhos\approx$ 20 -- 100 mK in experiments. Beyond this scale, the phase rotates to
take advantage of the tunneling energy. Therefore, we can interpret
the ground state as consisting of randomly polarized domains of size
$\ld$.  The total tunneling current in each domain is zero in the
ground state because the random field $\theta_0$ gives a current density $j_t$
[Eq.~\eqref{eq:tcurrent}] of random sign within each domain of
constant $\eta$.

However, if we drive the system by injecting currents phase twists
enter from the contacts, leading to configurations with non-zero
tunneling currents.  Injected counterflow current decays into the bulk
\emph{via} tunneling, \emph{i.e.}, recombination of excitons. This
tunneling current is supplied by rotating the phase of the domains
near the contacts, leaving domains in the bulk in their ground
state. The maximum coherent tunneling current that can be supported by
each domain is given by\cite{Eastham2010}:
\begin{equation}
\Id =e\rhos /\hbar\,.
\label{eq:domaincurrent}
\end{equation} 
If all the domains near the contact are rotated to supply this maximum
tunneling, then we see a uniform tunneling current $j_t$ until all
the injected counterflow current has decayed by tunneling.  Then both
the tunneling and counterflow currents are zero in the bulk.  This can
be described by the continuity equation for the current:
\begin{equation}
\ld^2\, \mbox{${\bm{\nabla}}\cdot\jcf$} = \pm \Id \,.
\label{continuity}
\end{equation}
where $\ld^2$ is the size of the domain and $\Id$ is the tunneling
current across the layers in the domain. The sign of $\Id$ is
determined by what is necessary to reduce the counterflow current. 
In a one-dimensional
geometry, this gives a counterflow current profile that decays
linearly into the bulk from the edge. For instance, for an injected
current of $-j_i$ at $x=0$, the current profile in the $x$-direction 
is
\begin{equation}
j_{\text{cf}}(x) = 
 \begin{cases}  
   -j_i+ \Id x/\ld^2\quad & 0<x<x_0\\
   0 & x>x_0\,.
  \end{cases}
\label{eq:profileinject}
\end{equation}
where the point $x_0 = j_i \ld^2/\Id$ marks the boundary of the region
from the sample edge where coherent tunneling occurs to reduce the
counterflow current. This current profile is a critical state similar
to the Bean critical state in superconductors\ \cite{Tinkham1996}. In
this state, there is a region near the contact over which the density
gradient in the soliton train (introduced by the injected current)
balances the pinning force arising from the tunneling.

\section{Susceptibility with disorder: zero-temperature limit}
\label{sec:zerot}

We now consider the response of the disordered system to an in-plane
field at zero temperature.  We will see that the disordered system has
a different diamagnetic response from the clean system.  It does not
have a commensurate-incommensurate transition controlled by an
intrinsic lengthscale $\lambda_J$ [see Eq.~\eqref{eq:b0}]. Instead, we
find a saturation phenomenon for the diamagnetic moment at a field
$B_c$ [see Eq.~\eqref{eq:ipsus}] controlled by the properties
[Eq.~\eqref{eq:domainsize} and \eqref{eq:domaincurrent}] of the
phase-polarized domains, \emph{i.e.}, the disorder-pinning of the
applied flux. This is followed by a depinning of the polarized domains
and a strong suppression of the diamagnetic response at a higher field
$B_{c2} > B_0$ [Eq.~\eqref{eq:bc2}] controlled by the disorder
correlation length $\xi$. Such behavior is similar to that of the
mixed-state of a superconductor, but very different from that
previously predicted for exciton condensates.

We require the response of the superfluid phase $\eta$ to an in-plane
field $B$, corresponding to a non-zero $\bq$ in
Eq.~\eqref{heff-etagauge}.  To obtain this, we note that
Eq.~\eqref{etapendulum} for the phase $\eta$ does not depend on the
in-plane field. Alternatively, we can see that the superfluid phase is
coupled to the field in Eq. \eqref{heff-etagauge} as an integral over
$\bq.\bnabla\eta$, which can be written as a boundary term for
$\eta$. Thus the current distribution in a field can be related to
that in zero field with shifted boundary conditions.  This observation
allows us to apply the critical-state model\cite{Lee2011,Eastham2010}
in zero field, as reviewed above, to calculate the current profile in
non-zero field.

For definiteness, we consider a rectangular sample with dimensions
$L_x$ and $L_y$, with the in-plane field in the $y$-direction. Thus,
counterflow currents $\jcf(x)$ flow in the $x$-direction, and
tunneling currents $j_t(x)$ in the $z$-direction (see
Fig.~\ref{fig:bilayer}).  For an isolated bilayer, $\jcf(x)= 0$ at
$x=0$ and $L_x$.  We can decompose $\jcf$ into paramagnetic and
diamagnetic parts as in Eq.~\eqref{eq:cfcurrent}.  The paramagnetic
part, $\jp=e\rhos\bnabla\eta/\hbar$, obeys the boundary condition
$\jp.\xuvec|_{x=0}=\jp.\xuvec|_{x=L_x}=-j_d$. Since the phase $\eta$
obeys the same equation, Eq.~\eqref{etapendulum}, in the bulk
irrespective of an in-plane field, $\jp$ has the same profile as in a
system in \emph{zero} field (when $\jcf=\jp$) with a current of $-j_d$
flowing across the boundaries.  The full counterflow current $\jcf$ for
the isolated bilayer in a field is recovered simply by adding a
uniform $+j_d$ to this zero-field profile of $j_p(x)$.

\begin{figure}[hbt]
\includegraphics{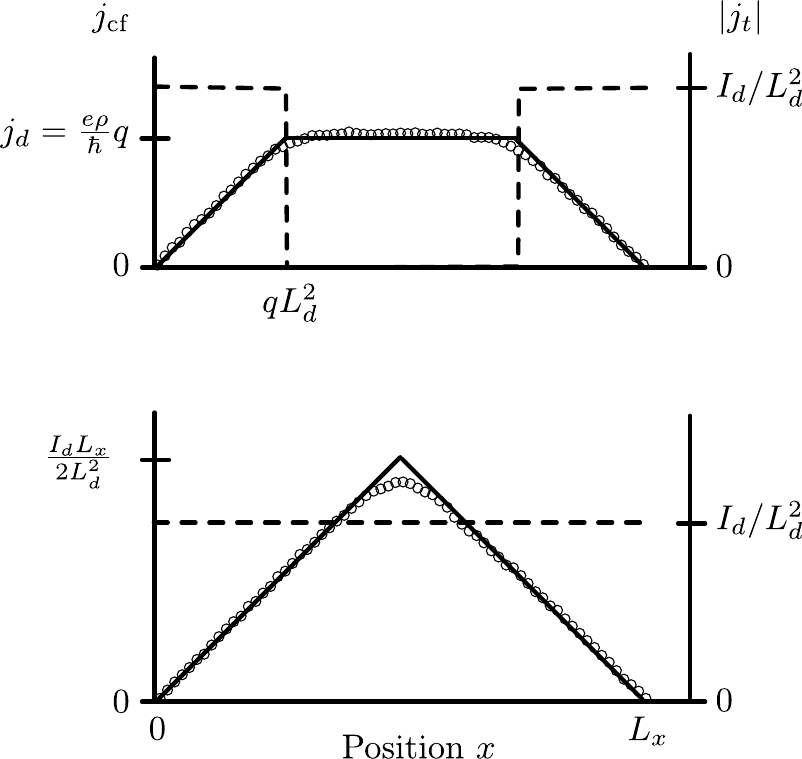}
\caption{Profiles of the counterflow (solid lines and circles, left
  axis) and tunneling (dashed, right axis) currents induced by
  applying an in-plane field $B$ to a rectangular bilayer
  (Fig.~\ref{fig:bilayer}), at zero temperature. The lines show
  results from the critical-state model, and the circles show
  numerical results from the microscopic Eq.~\eqref{etapendulum}.  The
  top (bottom) panel shows the profile below (above) the critical
  field $B_{c}$ [Eq.~\eqref{eq:critfield}]. The tunneling currents
  have opposite signs in the two halves of the sample (not shown). The
  simulation (circles) is a 200-site lattice model in one dimension
  with $ta^2/\rho_s=0.4$, where $a$ is the lattice spacing and
  $qa=6$ (top) and 12 (bottom). The disorder is uncorrelated from site
  to site ($\xi\approx a$).  }
\label{fig:critcprofs}
\end{figure}

In the critical-state model as discussed above, the counterflow
currents obey Eq.~\eqref{continuity} and so $j_p(x)$ decays linearly
in space from $-j_d$ at the edge to zero.  In other words, the profile
of the paramagnetic current $j_p(x)$ near the $x=0$ edge is 
given by Eq.~\eqref{eq:profileinject} with $j_i=j_d$.  The
distance $x_0$ which gives the width of the region where coherent
tunneling occurs is determined by the distance where the total
counterflow current vanishes: $j_p(x_0)+j_d = 0$. This criterion gives
a distance of 
\begin{equation}
x_0= j_d\ld^2/\Id=q\ld^2
\label{eq:tunnelwidth}
\end{equation}
from the edge of the sample.  The total current profile is shown in
Fig.~\ref{fig:critcprofs} (top). The figure also shows the tunneling
current density $|j_t|=\Id/\ld^2$ which is at its maximal value up to
the edge of the tunneling region at $q\ld^2$ from (either) edge of the
sample.  \derek{This is the start of the extended discussion to answer
  the referee.}  Beyond $x_0$, there is no paramagnetic contribution
to the counterflow current.

This current profile is similar to the one depicted in Fig.~\ref{fig:bilayer}. A diamagnetic
counterflow is generated by the in-plane field in the bulk of the
bilayer. However, an isolated bilayer must have zero counterflow current
at the edges.  In our critical-state model, this is
achieved by coherent tunneling \emph{via} phase-polarized domains. The
width of this tunneling region is determined by the size of the domain
$\ld$ and the coherent tunneling current that each could support
$\Id$. We see that a stronger field gives a higher diamagnetic
current and so more domains must be involved. In other words, the
size of the tunneling region increases with the in-plane field, as can
be seen from Eq.~\eqref{eq:tunnelwidth}:
$x_0 \propto q \propto B$. This
should be contrasted with the decay length $\lambda_J$ for the
commensurate state of the clean model which is an intrinsic scale that
does not vary with the field.

As the parallel field $B$ is increased, the diamagnetic current in the
bulk increases linearly with $B$. This requires a wider tunneling
region so that all the current can decay to zero at the edge. So, the
tunneling region increasingly penetrates the bulk.
This penetration is complete when the width of the region, $q\ld^2$, reaches
$L_x/2$. This saturation occurs when the field reaches a critical field of
\begin{equation} 
B_c=(\hbar/2e) (L_x/d \ld^2) = (\hbar/2e) (L_x \xi^2t^2/d\rhos^2)\,.
\label{eq:critfield}
\end{equation} 
Beyond this critical field, all the phase domains in the sample take
part in coherent tunneling with a constant magnitude for the tunneling
current density $j_t$, as depicted in Fig.~\ref{fig:critcprofs}
(bottom). The counterflow current, therefore, rises linearly from zero
from either edge, reaching a maximum at the center of the sample. This
saturated current profile stays the same for fields higher than
$B_c$. Note that, whereas the critical field $B_0$ in the clean limit
is a microscopic parameter independent of sample geometry, $B_c$ is
proportional to the length of the sample in a disordered system.

We have tested the critical-state model by comparing it with numerical
minimization of the energy [Eq.~\eqref{heff-etagauge}] for a
one-dimensional lattice model. To do this we add a dissipative
dynamical term $-\dot\eta$ to the right-hand side of
Eq.~\eqref{etapendulum} and numerically find the resulting
steady-state. We begin by finding the steady-state with the boundary
condition $d\eta/dx=0|_{x=0,L_x}$, corresponding to an isolated
bilayer in the absence of an in-plane field. We then slowly increase
the boundary condition so that $d\eta/dx|_{x=0,L_x}=-q$, corresponding
to applying the in-plane field. The counterflow supercurrent is then
obtained from the resulting phase profile by adding the diamagnetic
term, as described above. The resulting current profiles, averaged
over disorder realizations, are shown as the circles in
Fig.~\ref{fig:critcprofs}, and are seen to agree closely with the
critical-state model.


We will now discuss the magnetic moment generated by the diamagnetic
response of the bilayer.  Integrating the current profile gives the
moment $\Mpara$:
\begin{equation}
  \frac{\mu_0\Mpara}{L_x L_y d} =
 \begin{cases}  
   \chi_0 B\left(1-\frac{B}{2 B_c}\right) & B<B_c \\
   \chi_0 B_c/2 & B\ge B_c\,.
  \end{cases}
\label{eq:ipsus}
\end{equation} 
We see that the moment no longer rises linearly with $B$ with the full
diamagnetic susceptibility $\chi_0$ of the clean system.  This is
because, as seen in Fig.~\ref{fig:critcprofs}, the fraction of the
sample with the full diamagnetic current $j_d$ is continuously reduced
as $B$ (and hence $q$) increases. 
In fact, the total moment saturates at the critical
field $B_c$ to the value of
\begin{equation}
\Mparamax = 
\frac{e\rhos L_x}{4\hbar\ld^2} L_x L_y d\,.
\label{eq:maxmoment}
\end{equation}
The analysis above applies if the field and supercurrents are
sufficiently small that their presence does not modify the pinning
length $\ld$, \emph{i.e.}, the supercurrents do not depin the
vortices. This is the case if $q<1/\xi$, the inverse disorder correlation
length, 
so that the additional phase
winding at wavevector $q$ in Eq.~(\ref{heff-thetagauge}) does not
affect the tunneling energy. For larger fields however, the tunnel
currents will oscillate within each correlation area, rapidly
suppressing the net tunneling at larger scales.  The current $j_p$, which
is $j_d$ at the boundary, will then be approximately uniform, and
$\jcf$ and $j_{t}$ will be small at long wavelengths. Thus we expect
the susceptibility to fall quickly for fields beyond 
\begin{equation} 
B_{c2}\sim \frac{\hbar}{ed\xi}\sim 0.5 \mathrm{T}.
\label{eq:bc2}
\end{equation} This can be
regarded as a depinning field where the phase is no longer pinned by
the random field but is determined by the diamagnetic wavevector $q$.
A corresponding suppression\cite{Eastham2010} of the current in
interlayer tunneling experiments is expected, and observed, at such
fields.  Since $B_c/B_{c2} = L_x\xi/2\ld^2 <1$ for experimental
samples with $L_x \sim$ 1 mm, the saturation at $B_c$ occurs before
this suppression at $B_{c2}$. Note that in the lattice model discussed
above, the disorder is uncorrelated between sites, so that the
numerical simulations do not capture the depinning effect at $B_{c2}$.


\section{Flux creep and dissipation at low temperatures}
\label{sec:finitet}

Let us now consider thermal fluctuations over the energy barriers
which pin the winding of the superfluid phase. This allows the
supercurrents to relax to the equilibrium state with $\jcf=0$ and
$j_{t}=0$. The diamagnetic moments induced by an in-plane field will
decay as phase slips enter the system. This gives rise to resistivity
in transport measurements\cite{Huse2005}. We will estimate the size of
these effects, applying the conventional flux-creep model of
superconductors\cite{ANDERSON1964}.  We neglect the possibility of
vortex-glass states\cite{Nattermann2000}, which may further suppress
the dissipation at low temperatures.

For sufficiently low temperatures $T$, the dynamics of the merons
controlling the diamagnetic moment will involve thermal fluctuations
of the phase domains of size $\ld$ discussed above.  As in the
Anderson-Kim theory\cite{ANDERSON1964}, uncorrelated vortex motion
should be irrelevant at low temperatures and small bias, because the
vortex separation $\xi \ll \ld$, so the energy barriers are larger. A
thermal fluctuation in which the phase of a domain changes by
approximately $2 \pi$ corresponds to a phase slip moving a distance
$\ld$. This will occur at a frequency $\omega_0 e^{-U(j)/kT}$, where
$\omega_0$ is an attempt frequency and $U(j)$ is an energy barrier. Such
phase slip dynamics implies an interlayer voltage according to the
Josephson relation $V=\hbar \dot\theta/e$. For a typical domain size
of $\ld$, this gives a layer-antisymmetric in-plane electric field
\begin{equation}
  E=E_{\rm top}-E_{\rm bottom}\approx
  \frac{\hbar\omega_0}{e \ld}e^{-U(j)/kT}.
\label{eq:efields}
\end{equation}
At zero current, the typical energy barrier $U(j)$ will be the energy
of a domain, which is the stiffness $\rhos$
irrespective of size\cite{Eastham2010} in two dimensions.  
The barrier will vanish at the current scale associated with the
domain, 
$j_c=\Id/\ld \approx e\rhos/(\hbar \ld)$. 
The linear interpolation, $U(j)=\rhos(1-j/j_c)$,
corresponds to the Anderson-Kim
model. Inserting this form in
Eq.~\eqref{eq:efields}, we obtain for the ohmic regime 
a sheet resistance for counterflow currents of
\begin{equation} 
R_s \approx
  (\hbar^2\omega_0/e^2 kT) e^{- \rhos/kT}.\qquad (kT \ll \rhos)
\label{eq:sheetres}
\end{equation}
The in-plane magnetic moment, and the long-wavelength
counterflow supercurrent, relaxes in time $t$ as\cite{Gurevich1993}
\begin{equation} 
\Mpara(t)/M_{\parallel,T=0} \approx 1-(kT/\rhos)\ln (\omega_0t),
\label{eq:momentrel}
\end{equation} where $M_{\parallel,T=0}$ is the $T=0$ moment
from Eq.~\eqref{eq:ipsus}. 
 
It is possible to extend this argument into the dissipative regime at
high bias or temperature, where the barriers become irrelevant. The
simplest assumption would be that the domain rotates every attempt
time $1/\omega_0$ when the current density is $j_c$. This gives a
resistivity in the flux-flow regime of
\begin{equation} 
R^{*}_s \approx \frac{\hbar^2\omega_0}{e^2 \rho_s}, 
\label{eq:flowform} 
\end{equation} 
so that we can rewrite the flux-creep form as 
\begin{equation}
 R_s \approx R_s^*  e^{-\rhos/kT}.
\label{eq:creepform}
\end{equation} 
However, details of the dissipation are likely to change between these
two regimes and other parameters may enter $R_s^\ast$, so that the
prefactor in Eq.~(\ref{eq:creepform}) will not be exactly the
resistivity in the dissipative regime.  Provided that $R_s^\ast$ is
not fitted too far from the flux-creep regime, this is unimportant in
practice because the exponential factor in general dominates in
Eq.~(\ref{eq:creepform}).  Over a wider range of temperatures and
currents, effects such as the activation of quasiparticles may appear,
leading to additional exponential factors in the dissipation.

Reliable calculations of $\omega_0$ are very difficult, and irrelevant
in practice in the flux-creep form,
Eq.~(\ref{eq:creepform}). Nonetheless, an estimate may be obtained by
considering the dynamical equation for the
phase,\cite{Fogler2001,Wen1993}
\begin{equation}
  \nabla^2 \eta -
  \frac{1}{\lambda^2_J}\sin(\eta 
   +\theta_0) \\ 
  =\frac{\hbar^2}{e^2 \rhos} \left[c\ddot
  \eta - \frac{1}{\rho_z} \dot \eta -
  \frac{1}{\rho_{xx}} \nabla^2\dot \eta\right]
\label{eq:dynmics}
\end{equation}
where $\rho_{xx}$ ($\rho_z$) is the resistivity of in-plane
(tunneling) quasiparticle currents, and $c$ is the interlayer
capacitance per unit area. This gives several candidates for the
frequency scale at length scale $\ld$.
Since the resistivities are activated, we
expect the first (inertial) term in Eq.~(\ref{eq:dynmics}) to control the
dynamics at sufficiently low temperatures: $\omega_0=
(e/\hbar\ld)(\rhos/c)^{1/2}$.

\section{Discussion}
\label{sec:discuss}

We have examined the diamagnetic response of an isolated bilayer due
to counterflow superfluidity. We obtained an in-plane diamagnetic
moment which saturates at a critical value $B_c$.  Our theory also
predicts a field scale $B_{c2}$ for the suppression of the in-plane
diamagnetic response. There would be no in-plane susceptibility in a
field-cooled bilayer, even below $B_{c2}$, allowing a separation of
the superfluid signal.  We will now discuss briefly the magnitude and
timescales of these effects using realistic parameters for current
experiments.

From Eqs.~\eqref{eq:momentrel} and \eqref{eq:creepform}, we see that
the diamagnetic moment is long-lived, and the dissipation small, when
$kT\ll \rhos$. In previous work\cite{Eastham2010}, we estimated that
$\rhos\approx 20\,\mathrm{mK}$ in current experiments. This small
value arises from the finite interlayer separation, and the reduced
area of the sample containing superfluid in the coherence network
picture. Using this estimate as well as a similar area reduction for
the capacitance, we find for the attempt frequency $\omega_0 \approx$
300MHz. Since the superfluid stiffness $\rhos$ is around the lowest
temperatures achieved, we expect from Eq.~\eqref{eq:momentrel} that
the moments relax rapidly.  Similarly, counterflow dissipation due to
flux motion would be significant with $R_s^\ast \approx $ 1 -- 10 Ohm.

Activated forms similar to Eq.~\eqref{eq:creepform} have previously
been obtained for the residual counterflow resistivity based on
hopping of in-plane vortices (merons) within the coherence
network\cite{Fertig2005}. In contrast, here we have obtained this form
from the motion of phase-polarized domains, which corresponds to a
large-scale collective motion of the merons. However, we expect the
energy barriers for vortex motion in two dimensions to be the
superfluid stiffness up to logarithmic factors, so long as the
possibility of a vortex glass is excluded. Thus this form is not
dependent on the precise model of vortex motion. We note that
experiments on hole bilayers\cite{Tutuc2004} see this activated
behavior in the counterflow resistance with a value of 20 Ohms at 30
mK with an activated form that agrees with Eq.~\eqref{eq:sheetres} if
we use $\rhos\approx 100$ mK.  Electron bilayers\cite{Kellogg2004} fit
an approximately activated form with similar resistance values. A
complete theory of dissipation and transport is beyond the scope of
this work. Other mechanisms of dissipation will also play a role, such
as the thermal excitation of vortices\cite{Hyart2011} as well as
fermionic quasiparticles with an energy gap of the order of
$\rho_s$. It suffices to point out that the contribution from flux
motion is not negligible in current samples. However the activated
form of Eq.~\eqref{eq:creepform} suggests that a modest increase in
$\rho_s$, achievable by reducing the interlayer separation, would
allow the study of the nearly dissipationless behavior of the
flux-creep regime.

We now turn to the feasibility of measuring the diamagnetic response
of a bilayer at low temperatures. To do this, we compare the maximum
moment $\Mparamax$ [Eq.~\eqref{eq:maxmoment}] with the scale of the
perpendicular moment, $\Mperp = L_x L_y e\omega_c/2$, for an electron
gas in the integer quantum Hall regime with cyclotron energy
$\hbar\omega_c \approx$ 50K in this system. Using our previous
estimates $\rho_s\approx 20\,\mathrm{mK}$, and $\ld\approx
100\,\mu\mathrm{m}$, we obtain $\Mparamax/\Mperp =
(\rhos/2\hbar\omega_c )(L_x d/\ld^2)\sim L_x/(2000\, {\rm m})$ when
$d= 28$ nm.  This smallness of the effect, in comparison to the
conventional magnetic moment, makes this very challenging to measure
using torque magnetometry. This can also be inferred by noting that in
our theory the critical currents in tunneling experiments are the
maximum diamagnetic currents, and the former are experimentally in the
nano-ampere range.


However, we can exploit our understanding of the disordered isolated
bilayer to consider how this diamagnetic response can be increased. In
particular, we note that $\Mparamax$ per unit volume increases with
$L_x$ and with the number of polarized domains in the sample.
Note that the interlayer tunneling can be increased by many orders of magnitude compared
with the current samples by reducing the tunnel barrier.  Increasing
the tunneling strength reduces the domain size $\ld$. Also, the narrower
barriers possible with stronger tunneling allow a larger
$\rho_s$. Our approach holds up to the point where the domain size 
$\ld$ reaches the disorder lengthscale, $\xi \sim 100\, \mathrm{nm}$,
and the maximum field $B_c$ just reaches the depinning field
$B_{c2}$. At this point, we obtain a significant magnetic moment
$\Mparamax/\Mperp \gtrsim L_x/(1\, {\rm mm})$. Thus the diamagnetism
of the exciton superfluid should be evident in samples with strong
interlayer tunneling.
 
Although in the weak-tunneling samples the moments are small, the
pinning and dynamics of the in-plane flux can still be probed in
transport experiments. One consequence of the pinning picture is that
the critical current in a tunneling experiment should not be affected
by in-plane fields smaller than $B_{c2}$, in contrast to the clean
limit where the critical current is suppressed at fields\cite{Fil2010}
$B_0<B_{c2}$.  This is because in the pinning picture it is the
\emph{gradients} of the in-plane flux density which drive the flux
through the disorder, and while the injected currents in the tunneling
geometry impose such gradients, a uniform in-plane field does not. It
would also be possible to study the dynamics of the in-plane flux in
tunneling experiments. An interlayer voltage at one end of the sample
introduces in-plane flux at a given rate, and measuring the interlayer
voltages and currents reveals the subsequent motion\
\cite{Gurevich1993} of this flux. The logarithmic relaxation of the
flux-creep regime can lead to hysteresis, which is seen in tunneling
experiments\ \cite{Tiemann2009}.

An interesting extension of our work would be to consider samples in
the Corbino disk geometry. In such a geometry, a radial magnetic field
could induce circulating counterflow supercurrents, similar to the
induction of linear counterflow supercurrents by an in-plane field in
the open geometry of Fig.~\ref{fig:bilayer}. In the open geometry,
tunneling is necessary to close the current loop, which we have seen
gives rise to a maximum magnetic moment at the field $B_c$. In a
Corbino disk, tunneling is not needed to close a circulating current
loop, so there may be no saturation effect at $B_c$ and the full
magnetic response may persist up to $B_{c2}$.  However, there will
still be radial variations in the densities of applied flux and
diamagnetic current, and maintaining the diamagnetic current loop
requires that these variations are pinned. Thus it seems likely that,
even in this geometry, there is a maximum magnetic moment determined
by the sample dimensions and a pinning length. The relevant pinning
length, however, may be determined by parameters other than the
tunneling strength.

\section{Conclusions}
\label{sec:conclusions}

In summary, we have considered the diamagnetic response of a bilayer
exciton superfluid to an in-plane magnetic field, in the presence of
in-plane vortices nucleated by charge disorder. We argue that at low
temperatures, changes in the in-plane magnetic field induce circulating
diamagnetic currents and hence persistent diamagnetic moments. The
maximum moments which can be induced are determined by the pinning of
the in-plane flux by the disorder, which involves a characteristic
lengthscale related to the tunneling. At finite temperatures, thermal
motion of the in-plane flux will lead to transport resistivities and
cause the diamagnetic moments to decay. In current samples, we find that
the maximum moments are small, but samples with stronger tunneling
could allow persistent exciton supercurrents to be probed by torque
magnetometry. Such experiments would be analogous to the torsional
oscillator experiments that are the definitive measure of superfluid
fraction in helium.

Finally, we note that very strong tunneling, or very large stiffness,
may even be able to prevent the disorder nucleating
vortices\cite{Eastham2009a}, recovering the clean limit in which the
magnetic response is also expected to be
measurable\cite{Hanna2001}. In this case the phase diagram of the
bilayer would then closely parallel that of a type-II superconductor,
with both a mixed state (as described here) and a clean state,
experimentally distinguishable by the presence of hysteresis.

\acknowledgments We acknowledge support from Science Foundation
Ireland (SFI/09/SIRG/I1592) (PRE) and EPSRC EP/F032773/1 (NRC), and we thank J. P. Eisenstein for helpful discussions.


\end{document}